\documentclass[aps,prd,twocolumn,nofootinbib,showpacs]{revtex4-1}

\usepackage{amsfonts}
\usepackage{amsmath}
\usepackage{amssymb}
\usepackage{bm}
\usepackage{dcolumn}
\usepackage{graphicx}
\usepackage{graphics}
\usepackage[latin1]{inputenc}
\usepackage{latexsym}
\usepackage{rotating}
\usepackage{hyperref}
\usepackage[all]{hypcap} 
\usepackage{xspace} 
\usepackage[usenames,dvipsnames]{color}
\usepackage{mathrsfs}
\usepackage{subfigure}
\usepackage{amsfonts}
\usepackage{booktabs}
\usepackage{siunitx}

\usepackage{outlines}
\usepackage{enumitem}
\setenumerate[1]{label=\Roman*.}
\setenumerate[2]{label=\Alph*.}
\setenumerate[3]{label=\roman*.}
\setenumerate[4]{label=\alph*.}

\definecolor {darkgreen}{rgb}{0.2,0.7,0.2}

\newcommand{\be}{\begin{equation}}
\newcommand{\ee}{\end{equation}}
\newcommand\ba{\begin{eqnarray}}
\newcommand\bse{\begin{subequations}}
\newcommand\ea{\end{eqnarray}}
\newcommand\ese{\end{subequations}}

\newcommand{\order}[1]{\mathcal{O}\left(#1\right)}
\newcommand{\avg}[1]{\left \langle #1 \right \rangle}

\newcommand{\pd}{\partial}

\newcommand{\GAE}{G_{\AE{}}}

\newcommand{\TT}{{\mbox{\tiny TT}}}

\newcommand{\vt}{{\text{v}_{\tiny T}}}
\newcommand{\vv}{{\text{v}_{\tiny V}}}
\newcommand{\vs}{{\text{v}_{\tiny S}}}

\begin{document}
\title{Angular Momentum Loss for a Binary System in Einstein-\AE{}ther Theory}

\author{Alexander Saffer}
\affiliation{eXtreme Gravity Institute, Department of Physics, Montana State University, Bozeman, MT 59717, USA.}

\author{Nicol\'as Yunes}
\affiliation{eXtreme Gravity Institute, Department of Physics, Montana State University, Bozeman, MT 59717, USA.}

\date{\today}

\begin{abstract}
The recent gravitational wave observations provide insight into the extreme gravity regime of coalescing binaries, where gravity is strong, dynamical and non-linear.
The interpretation of these observations relies on the comparison of the data to a gravitational wave model, which in turn depends on the orbital evolution of the binary, and in particular on its orbital energy and angular momentum decay.  
In this paper, we calculate the latter in the inspiral of a non-spinning compact binary system within Einstein-\AE{}ther theory. 
From the theory's gravitational wave stress energy tensor and a balance law, we compute the angular momentum decay both as a function of the fields in the theory and as a function of the multipole moments of the binary. 
We then specialize to a Keplerian parameterization of the orbit to express the angular momentum decay as a function of the binary's orbital elements.  
We conclude by combining this with the orbital energy decay to find expressions for the decay of the semi-major axis and the orbital eccentricity of the binary. 
We find that these rates of decay are typically faster in Einstein-\AE{}ther theory than in General Relativity due to the presence of dipole radiation. 
Such modifications will imprint onto the chirp rate of gravitational waves, leaving a signature of Einstein-\AE{}ther theory that if absent in the data could be used to stringently constrain it.
\end{abstract}

\maketitle
\allowdisplaybreaks[4]

\section{Introduction}
\label{Intro}

Since the first detection of gravitational waves (GW) by the advanced Laser Interferometer Gravitational-Wave Observatory (aLIGO), gravitational wave astronomy has moved to the forefront of scientific study~\cite{Abbott:2016blz,Abbott:2016nmj,Abbott:2017vtc,Abbott:2017gyy,Abbott:2017oio,TheLIGOScientific:2017qsa}. These observations have confirmed the existence of gravitational radiation emanating from coalescing compact binaries. The GWs emitted depend sensitively on the orbital evolution of such binaries, and thus, their observation is a prime target for new tests in the \textit{extreme gravity} regime, where the gravitational interaction is simultaneously strong, dynamical and non-linear. 

Einstein-\AE{}ther theory (\AE{})~\cite{Jacobson:2000xp} is an ideal model to test with GWs because it encompasses one of the most general ways to break one of General Relativity's (GR) fundamental pillars: local Lorentz invariance. This symmetry is broken through a unit timelike vector field, called the \ae{}ther field, which couples non-minimally to the metric tensor and induces the activation and propagation of scalar, vectorial, and tensorial metric perturbations in compact binary systems~\cite{Jacobson:2007fh,Foster:2005dk,Hansen:2014ewa}. Recent gravitational wave observations have stringently constrained the speed of propagation of tensor modes~\cite{GBM:2017lvd}, but they could not say anything about the other modes, which are only constrained by Solar System and binary pulsar observations much more weakly~\cite{Jacobson:2007fh,Foster:2005dk,Yagi:2013ava}.    

Detailed GW tests of \AE{}-theory require the construction of GW models in the theory, which in turn requires an understanding of coalescing binaries. During the inspiral, their trajectory can be modeled as a sequence of osculating Keplerian orbits, with adiabatically changing semi-major axis and orbital eccentricity, within the post-Newtonian (PN) approximation~\cite{Peters:1963ux,Peters:1964zz}. These adiabatic changes are controlled by the rate at which the binary loses orbital energy and angular momentum. In \AE{}-theory, these rates are modified due to the activation of scalar and vectorial tensor perturbations that propagate away from the binary and carry energy and angular momentum away with them~\cite{Foster:2006az,Foster:2007gr}. This, in turn, modifies the chirping rate of GWs, leaving a signature in the GW phase that could be used to constrain the theory if it is found to be absent from the data.  

We here study the energy and the angular momentum loss rate of a binary system in \AE{}-theory to, for the first time, derive the rate at which the semi-major axis and the orbital eccentricity decays due to GW emission. We first compute the GW, stress-energy pseudo-tensor (SET) by expanding the field equations to second order in perturbations about a flat background, a method developed by Isaacson~\cite{Isaacson:1967zz,Isaacson:1968zza}. We then use this SET to compute the energy and the angular momentum flux carried by all propagating modes in terms of derivatives of these fields~\cite{Landau:1982dva}. With this at hand, we use a multipolar expansion of the fields to calculate these fluxes in terms of derivatives of multipole moments. Lastly, we use a Keplerian parametrization to present the energy and angular momentum flux as a function of the orbital parameters of the binary. In doing so, we verify the energy flux calculations of~\cite{Yagi:2013ava}, extending them to include the angular momentum flux for the first time. This last result allows us to calculate the rate of decay of the semi-major axis and the orbital eccentricity of the binary. 

\begin{figure}
	\centering
	\vspace{0.4cm}
	\includegraphics[width=0.49\textwidth]{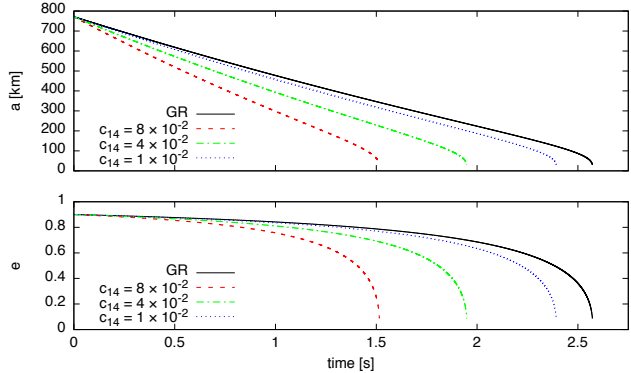}
	\caption{Temporal evolution of the semi-major axis (a) and the orbital eccentricity (e) of a 1.4$M_\odot$-2$M_\odot$ neutron star binary due to the modified decay rates of the binary's orbital energy and angular momentum. To construct this figure, we modeled the sensitivities with the weak-field expansion of~\cite{Foster:2006az,Yagi:2013ava}, and we used an initial orbital frequency of 10 Hz and an initial orbital eccentricity of 0.9. Observe that the semi-major axis and the orbital eccentricity decay faster than in GR. }
	\label{fig:changes}
\end{figure}

We find that generically the semi-major axis and the eccentricity decay rate is faster in \AE{}-theory than in GR when dipole emission is present. Figure~\ref*{fig:changes} shows these evolutions for different choices of the $c_{14}$ combination of coupling constants of the theory. All other coupling constants are chosen such that all propagating modes travel at the speed of light, thus satisfying stability and Cherenkov constraints~\cite{Elliott:2005va}, and leaving $c_{14}$ as the only free parameter of the theory that allows us to continuously approach the GR limit~\cite{Barausse:2015frm}. Observe that the evolution of the semi-major axis and the orbital eccentricity becomes faster than in GR (the $c_{14}=0$ case) as the combination of coupling constants $c_{14}$ is increased. Such an increased rate of decay will imprint onto the chirping rate of the emitted GWs, which could be constrained in the future with GW observations. 

The structure of this paper is as follows. 
Section~\ref{sec:EinsteinAether} presents the basics of \AE{}-theory, including how we decomposes the fields of the theory into different propagation modes (Sec.~\ref{sec:FieldDecompositions}).
Section~\ref{sec:SETandLDOT} explains how to derive the SET and how to calculate the angular momentum loss rate.  
Section~\ref{sec:BinarySection-AE} specializes the previous results to a binary system in a Keplerian parametrization, calculating the rate of semi-major axis and orbital eccentricity decay. 
Section~\ref{sec:Conclusions} concludes and discusses future implications. 
In the remainder of this paper, we use the standard $\left(-,+,+,+\right)$ metric signature as described in~\cite{Misner:1974qy}, as well as units in which $c=1$.

\section{Einstein-\AE{}ther Theory}
\label{sec:EinsteinAether}
We begin with a brief introduction to \AE{}-theory, which was developed by Jacobson and Mattingly in 2001 by coupling a Lorentz violating vector field to GR~\cite{Jacobson:2000xp}.  We here follow mostly the presentation~\cite{Foster:2006az}, where the full action of the theory can be written as a sum of a gravitational term and a matter term
\begin{equation}
\label{eq:full-action}
S = S_{\AE{}} + \sum_{A} S_{A}\,,
\end{equation}
for $A$ bodies in the system.

\subsection{Gravitational Action}

The gravitational action in $\AE$-theory is
\begin{align}
\label{eq:EA_Action}
S_{\AE{}} &= \frac{1}{16 \pi G_{\AE}} \int d^4x \sqrt{-g}\left(R \right. \nonumber \\
&\left. - {K^{\alpha \beta}}_{\gamma \epsilon} \, \nabla_\alpha u^\gamma \nabla_\beta u^\epsilon + \lambda \left(u^\alpha u_\alpha +1\right)\right)\,,
\end{align}
where 
\begin{align}
\label{eq:EA_Keqn}
{K^{\alpha \beta}}_{\gamma \epsilon} = c_1g^{\alpha \beta}g_{\gamma \epsilon}+c_2{\delta^\alpha}_\gamma{\delta^\beta}_\epsilon + c_3{\delta^\alpha}_\epsilon{\delta^\beta}_\gamma - c_4u^\alpha u^\beta g_{\gamma \epsilon}\,,
\end{align}
$R$ is the Ricci tensor associated with the metric $g_{\alpha \beta}$, and $u^\alpha$ is the unit timelike \ae{}ther field constrained via a Lagrange multiplier $\lambda$.  The coupling constants $c_i$ control the various ways in which the \ae{}ther field couples to the metric.  For future convenience, we introduce the following combinations of $c_i$:
\begin{subequations}
	\begin{align}
		c_{\pm} &= c_1 \pm c_3 \,,\\
		c_{14} &= c_1 + c_4 \,,\\
		c_{123} &= c_1 + c_2 + c_3\,.
	\end{align}
\end{subequations}
The term $G_{\AE}$ in Eq.~\eqref{eq:EA_Action} is the modified gravitational constant which is related to the Newtonian one through the relation~\cite{Carroll:2004ai,Yagi:2013ava}
\begin{equation}
	G_N = \frac{2}{2-c_{14}}G_{\AE{}}\,.
\end{equation}

\subsection{Matter Action}

For compact objects modeled as effective point-particles, the matter action of the $A^{\text{th}}$ particle is given by~\cite{Foster:2007gr}
\begin{subequations}
	\begin{align}
		\label{eq:MatterAction1}
		S_A &= - \int d\tau_A \, \tilde{m}_A(\gamma_A)\, \\
		\label{eq:MatterAction2}
		&= -\tilde{m}_A \int d\tau_A \left[+\sigma_A\left(\gamma_A+1\right) + \frac{1}{2} \sigma_A^\prime \left(\gamma_A+1\right)^2 + \cdots\right]\,,
	\end{align}
\end{subequations}
where $d\tau_A$ is the proper time along the particle's world-line, $\gamma_A = -u_\alpha v_A^\alpha$ with $v_\alpha$ the particle's four-velocity, and $\tilde{m}_A$ is the bare mass, which is related to the particle's active gravitational mass via 
\begin{equation}
\label{eq:ActiveGravMass}
	m_A = \left(1+\sigma_A\right)\tilde{m}_A\,.
\end{equation}
We define the $\sigma_A$ and $\sigma_A^\prime$ charges in Eq.~\eqref{eq:MatterAction2} as
\begin{subequations}
	\begin{align}
		\sigma_A &= - \frac{d\,\ln(\tilde{m}_A)}{d \ln(\gamma_A)}|_{\gamma_A = 1}\,, \\
		\sigma_A^\prime &= \sigma_A + \sigma_A^2 + \frac{d^2\,\ln(\tilde{m}_A)}{d \left(\ln(\gamma_A)\right)}|_{\gamma_A = 1}\,,
	\end{align}
\end{subequations}
and relate them to a rescaled sensitivity parameter by~\cite{Yagi:2013ava}
\begin{equation}
\label{eq:sDef}
	s_A = \frac{\sigma_A}{1+\sigma_A}\,.
\end{equation}

\subsection{Field Equations}
\label{sec:FieldEquations}

The variation of the full action in Eq.~\eqref{eq:full-action} with respect to the metric yields the \AE{} modification to the Einstein equations~\cite{Carroll:2004st}
\begin{equation}
\label{eq:FieldEquations}
	G_{\alpha \beta} - S_{\alpha \beta} = 8 \pi \, T_{\alpha \beta}\,,
\end{equation}
where $G_{\alpha \beta}$ is the Einstein tensor, and $S_{\alpha \beta}$ is given by
\begin{align}
\label{eq:S_Tensor}
S_{\alpha \beta} &= \nabla_\gamma \left({K^\gamma}_{(\alpha} u_{\beta)} + K_{(\alpha \beta)} u^\gamma - {K_{(\alpha}}^\gamma u_{\beta)}\right) \nonumber \\
& + c_1 \left(\nabla_\alpha u_\gamma \, \nabla_\beta u^\gamma - \nabla_\gamma u_\alpha \, \nabla^\gamma u_\beta\right) \nonumber \\
& + c_4 \left(u^\gamma  u^\delta \, \nabla_\gamma u_\alpha \, \nabla_\delta u_\beta\right) + \lambda \, u_\alpha u_\beta \nonumber \\
& - \frac{1}{2}g_{\alpha \beta}\left({K^\gamma}_\delta \, \nabla_\gamma u^\delta\right)\,.
\end{align}
with the contracted $K$-tensor
\begin{equation}
\label{eq:Contracted_K}
{K^\alpha}_\gamma = {K^{\alpha \beta}}_{\gamma \delta} \nabla_\beta u^\delta\,.
\end{equation}

The matter stress-energy tensor $T_{\alpha \beta}$ is defined via
\begin{equation}
	\label{eq:MatterSET}
	T_{\alpha \beta} = - \frac{2}{\sqrt{-g}}\frac{\partial S_m}{\partial g^{\alpha \beta}}\,,
\end{equation}
which can be calculated through the matter action in Eq.~\eqref{eq:MatterAction2}:
\begin{align}
	T^{\alpha \beta} &= \sum_A \tilde{m}_A \tilde{\delta}_A \left[\left(1+\sigma_A-\frac{1}{2}\sigma_A^\prime\left(\gamma_A^2-1\right)\right)v_A^\alpha v_A^\beta \right. \nonumber \\
	&\left. - 2 \left(\sigma_A + \sigma_A^\prime\left(\gamma_A+1\right)\right)u^{(\alpha}v_A^{\beta)}\right]\,,
\end{align}
where
\begin{equation}
	\tilde{\delta}_A = \frac{1}{v_A^0 \sqrt{-g}}\delta^3\left(\vec{x} - \vec{x}_A\right)\,
\end{equation}
is short hand for a renormalized Dirac delta function. 

The variation of the full action in Eq.~\eqref{eq:full-action} with respect to the \ae{}ther field $u^\alpha$ yields the \ae{}ther field equation
\begin{equation}
	\label{eq:AEtherFE}
	\nabla_\alpha K^{\alpha \beta} = \lambda\,u^\beta + c_4\left(u^\alpha \,\nabla_\alpha u_\gamma\right) \nabla^\beta u^\gamma + 8\pi G_{\AE{}}\Upsilon^\beta\,,
\end{equation}
where
\begin{equation}
	\Upsilon^\alpha = - \sum_A \tilde{m}_A \tilde{\delta}_A \left[\sigma_A+\sigma_A^\prime \left(\gamma_A+1\right)\right]v_A^\alpha\,.
\end{equation}

In addition to the modified Einstein equations and the \ae{}ther field equation, \AE{}-theory also possesses the timelike constraint
\begin{equation}
\label{eq:TimelikeConstraint}
	g_{\alpha \beta} \, u^\alpha u^\beta = -1\,
\end{equation}
which can be derived by varying the full action in Eq.~\eqref{eq:full-action} with respect to the Lagrange multiplier $\lambda$.  With this we can solve Eq.~\eqref{eq:AEtherFE} for $\lambda$ to find
\begin{equation}
	\lambda = 8\pi G_{\AE{}}\Upsilon_\alpha \, u^\alpha+ c_4\left(u^\beta \nabla_\beta u_\gamma\right)\left(\nabla_\alpha u^\gamma\right)u^\alpha - \nabla_\beta K^\beta{}_\alpha\,u^\alpha\,.
\end{equation}

\subsection{Field Decomposition}
\label{sec:FieldDecompositions}

We conclude this section by carrying out an irreducible decomposition of the perturbations of all the fields about a fixed background, as done for example in~\cite{Foster:2006az}. We begin by expanding the metric
\begin{equation}
	g_{\alpha \beta} = \eta_{\alpha \beta} + h_{\alpha \beta}\,,
\end{equation}
about Minkowski spacetime $\eta_{\alpha \beta}$, with $|h_{\alpha \beta}| \ll |\eta_{\alpha \beta}|$\,, and expanding the \ae{}ther field
\begin{equation}
	u^\alpha = t^\alpha + \omega^\alpha\,,
\end{equation}
about $t^\alpha = (-1,0,0,0)$ a timelike background unit vector, with $|\omega^\alpha| \sim \order{h}$.  

Next, we perform two more decompositions. We begin with a $3+1$ decomposition of the metric perturbation about a spatial hypersurface normal to the timelike background vector
\begin{equation}
	h^{\alpha \beta} = t^\alpha t^\beta h_{00} + 2 \mathcal{P}^{(\alpha}_i t^{\beta)} h^{0i} + \mathcal{P}^{\alpha}_i \mathcal{P}^\beta_j h^{ij} \,,
	\end{equation}
where $\mathcal{P}^\alpha_i$ is the background spatial projector
	\begin{equation}
	\label{eq:SpatialProjector}
	\mathcal{P}^{\alpha \beta} = \eta^{\alpha \beta} + t^\alpha t^\beta\,.
\end{equation}
We then further decompose the $3+1$-decomposed metric tensor into transverse and longitudinal pieces
\begin{subequations}
	\label{eq:hDecomp}
	\begin{align}
		h^{0i} &= \gamma^i + \partial^i\,\gamma\,,\\
		h^{ij} &= \phi^{ij}_\TT + \frac{1}{2}\mathcal{F}_{ij}\left[f\right] + 2 \partial_{(i} \phi_{j)} + \partial_i \partial_j \phi \,,
	\end{align}
\end{subequations}
where $\phi^{ij}_\TT$ is a transverse-traceless spatial tensor, $\mathcal{F}_{ij}[f] \equiv \eta_{ij}\,F - \partial_i \, \partial_j f$ with $F=\partial_k \, \partial^k f$, and $\gamma^{i}$ and $\phi^{i}$ are longitudinal so that $\gamma^i{}_{,i}= \phi^i{}_{,i}=0$. We similarly decompose the \ae{}ther field into transverse and longitudinal modes through
\begin{equation}
	\omega^\alpha = t^\alpha \omega_0 + \mathcal{P}^\alpha_i \left(\nu^i + \partial^i \nu \right) \,, \\
\end{equation}
with $\nu^{i}$ longitudinal so that $\nu^i{}_{,i}=0$.  

We can simplify the above expressions somewhat as follows. First, we use the timelike constraint on the \ae{}ther field in Eq.~\eqref{eq:TimelikeConstraint} to find 
\begin{equation}
	\omega_0 = -\frac{h_{00}}{2}\,,
\end{equation}
to leading order in the metric perturbation. Next, we use the remaining gauge freedom to set 
\begin{subequations}
	\begin{align}
	\partial_i \omega^i &= 0 \,, \\
	\partial_i h^{0i}&= 0 \,, \\
	\partial_i \partial_{[j}{h_{k]}}^i &=0 \,,
	\end{align}
\end{subequations}
which grants us $\nu = \gamma = \phi_i =0$ upon simplification.  Through the evaluation of the field equations given in Eq.~\eqref{eq:FieldEquations} and~\eqref{eq:AEtherFE}, we are able to find relations which exist between the various remaining fields.  By looking at the tensor, vector, and scalar modes of the equations separately, we are able to obtain\footnote{For details on the calculation of these relations see~\cite{Foster:2006az,Saffer:2017ywl}.}
\begin{subequations}
	\begin{align}
		\gamma^i &= -c_{+} \, \nu^i \,, \\
		h_{00} &= \frac{1}{c_{14}}\,F\,,\\
		\partial_i \partial^i \phi &= - \frac{1+c_2}{c_{123}}\,F \,.
	\end{align}
\end{subequations}

With these simplifications at hand, one finds that \AE{}-theory only presents 3 non-vanishing perturbations: $\phi^\TT_{ij}$, $\nu^i$, and $F$.  These modes propagate as waves about the background spacetime, satisfying modified wave equations in vacuum of the form
\begin{subequations}
	\begin{align}
		\left(\frac{1}{\vt^2}\pd_0^2 - \pd^i\pd_i\right)\phi^\TT_{jk} &= 0\,,\\
		\left(\frac{1}{\vv^2}\pd_0^2 - \pd^i\pd_i\right)\nu^k &= 0\,,\\
		\left(\frac{1}{\vs^2}\pd_0^2 - \pd^i\pd_i\right)F &= 0\,,
	\end{align}
\end{subequations}
where the various propagation speeds are given by 
\begin{subequations}
	\begin{align}
	\vt^2 &= \frac{1}{1-c_+}\,, \\
	\vv^2 &= \frac{2\,c_1 - c_+c_-}{2 \, c_{14}\left(1-c_+\right)}\,,\\
	\vs^2 &= \frac{c_{123}\left(2-c_{14}\right)}{c_{14}\left(1-c_+\right)\left(2+3c_2+c_+\right)}\,.
	\end{align}
\end{subequations}
The recent coincident electromagnetic and GW observation of a neutron star merger~\cite{GBM:2017lvd} has placed a severe constraint on $\vt$, effectively forcing $c_{+} \lesssim 10^{-15}$, but this observation places no constraint on $\vv$ or on $\vs$. 

\section{Angular Momentum Loss Rate}
\label{sec:SETandLDOT}

In this section, we calculate the angular momentum loss rate in \AE{}-theory. We begin with a generic calculation
that leads to a result in terms of derivatives of the decomposed fields. We then specialize these results to an N-body
system and provide the angular momentum loss in terms of derivatives of the multipole moments of the system. 
In both cases, we also present previously-derived expressions for the energy loss for the sake of completeness.

\subsection{General Calculation}

The angular momentum loss can be calculated from~\cite{Landau:1982dva,PW}
\begin{equation}
\dot{L}^i \equiv \frac{1}{2} \epsilon^i{}_{jk}\dot{L}^{jk}\,,
\end{equation}
where $\epsilon^i{}_{jk}$ is the Levi-Civita symbol and where we have defined
\begin{equation}
\label{eq:LdotDefinition}
	\dot{L}^{ij} = -2\oint x^{[i}\,\Theta^{j]k} \, dS_k\,,
\end{equation}
with $\Theta^{ij}$ the spatial components of the GW SET.  The latter can be calculated in \AE{}-theory using the techniques outlined in~\cite{Isaacson:1967zz,Isaacson:1968zza,Saffer:2017ywl} by expanding the field equations to second order in the metric perturbation. The resulting expression cannot be simplified directly through Brill-Hartle averaging~\cite{PhysRev.135.B271}, because the integrand of $\dot{L}^{ij}$ is not $\Theta^{ij}$ but $x^{[i}\,\Theta^{j]k}$; one can of course simplify the latter through Bill-Hartle averaging, and we make use of this henceforth. 

The resulting expression can be simplified further through the \emph{shortwave approximation}~\cite{Misner:1974qy,PW}. This assumes the observer is in a region far removed from the source of  radiation, such that the observer's local radius of curvature $R_c$ is much larger than the characteristic wavelength of the gravitational radiation, $\lambda_c$.  This  allows us to use $\lambda_c/R_c \ll 1$ as an expansion parameter and simplify partial spatial derivatives on any field $\Phi$ via
\begin{equation}
	\label{eq:ShortwaveApprox}
	\partial_i \Phi = -\frac{N_i}{\text{v}_\Phi} \partial_\tau \Phi + \bar{\pd}_i \Phi\,,
\end{equation}
where $\text{v}_\Phi$ is the speed of the field in question and the special derivative $\bar{\pd}$ operates only on the spatial (typically radial) part of the field's argument. Therefore, terms proportional to  $(\bar{\pd} \partial_\tau \Phi)/\Phi$ are of $\mathcal{O}(\lambda_c/R_c)$.  

When we combine this approximation with the properties of the decomposed fields, we can further simplify the resulting expressions. For instance, we have that
\begin{subequations}
	\label{eq:ShortwaveConstraints}
	\begin{align}
	N_i \phi^{ij}_\TT = N_i \nu^i &= 0\,,\\
	N_i \pd_\tau \phi^{ij}_\TT = N_i \pd_\tau \nu^i &= 0\,,\\
	N_i \bar{\pd}_j \phi^{ik}_\TT &= -\frac{{\phi^{\TT k}}_{j}}{r}\,,\\
	\bar{\pd}_i N^j &= \frac{^{(3)}\mathcal{P}^j{}_i}{r}\,,
	\end{align}
\end{subequations}
where $^{(3)}\mathcal{P}_{ij} = \delta_{ij} - N_i\,N_j$ is a projector to a hypersurface orthogonal to the wave's propagation vector $N^{i}$, and $r$ is the distance to the source. Clearly the hypersurface orthogonal to $t^{\alpha}$ need not coincide with that orthogonal to $N^{i}$ and thus $^{(3)}\mathcal{P}_{ij} \neq \mathcal{P}_{ij}$.  We are required to utilize the second term in Eq.~\eqref{eq:ShortwaveApprox} in the $\dot{L}^{ij}$ calculation due to the presence of an extra radial term present in Eq.~\eqref{eq:LdotDefinition}.

After using all of these simplifications, the end result for $\dot{L}^{ij}$ is
\begin{equation}
\dot{L}^{ij} = \dot{L}^{ij}_{(T)} + \dot{L}^{ij}_{(V)} + \dot{L}^{ij}_{(S)}\,,  
\end{equation}
with 
\begin{subequations}
	\begin{align}
		\label{eq:FinalTensorLdot}
		\dot{L}^{ij}_{(T)} &= -\frac{1}{8 \pi \, \vt \, G_{\AE{}}} \int r^2 \left[\phi_\TT^{a[i} \left(\pd_\tau \phi_\TT^{j]}{}_a\right) \right. \nonumber \\ 
		& \left. - \frac{1}{2}\left(\pd_\tau \phi_{ab}^\TT\right)x^{[i} \bar{\pd}^{j]}\phi^{ab}_\TT\right] d\Omega\,, \\
		\label{eq:FinalVectorLdot}
		\dot{L}^{ij}_{(V)} &= -\frac{(1-c_+)(2c_1-c_+c_-)}{8 \pi \, \vv \,G_{\AE{}}} \int r^2 \left[\nu^{[i} \left(\pd_\tau \nu^{j]}\right) \right. \nonumber \\
		& \left. - \left(\pd_\tau \nu^a\right)x^{[i} \bar{\pd}^{j]}\nu_a\right] d\Omega\,,\\
		\label{eq:FinalScalarLdot}
		\dot{L}^{ij}_{(S)} &= -\frac{\left(2-c_{14}\right)}{32 \pi \, \vs \, c_{14} G_{\AE{}}} \int r^2 \left(\pd_\tau F\right)x^{[i} \bar{\pd}^{j]}F d\Omega \,.
	\end{align}
\end{subequations}
and the analogous result for the angular momentum loss is
\begin{equation}
\dot{L}^{i} = \dot{L}^{i}_{(T)} + \dot{L}^{i}_{(V)} + \dot{L}^{i}_{(S)}\,,  
\end{equation}
with 
\begin{subequations}
	\label{eq:AngularMomentomVector}
	\begin{align}
	\dot{L}^i_{(T)} &= -\frac{1}{16 \pi \, \vt \GAE} \int r^2 \, \epsilon^i{}_{jk} \left[\phi^{ja}_\TT \left(\pd_\tau \phi_\TT^k{}_a \right) \right. \nonumber \\
	& \left. -\frac{1}{2}x^j \left(\pd^k \phi^{ab}_\TT\right) \left(\pd_\tau \phi_{ab}^\TT\right)\right] d\Omega \,, \\
	\dot{L}^i_{(V)} &= -\frac{(1-c_+)(2c_1-c_+c_-)}{16 \pi \, \vv \GAE} \int r^2 \, \epsilon^i{}_{jk} \left[\nu^j \left(\pd_\tau \nu^k\right) \right. \nonumber \\
	& \left. - x^j \left(\pd^k \nu^a\right)\left(\pd_\tau \nu_a\right)\right] d\Omega\,, \\
	\dot{L}^i_{(S)} & = -\frac{\left(2-c_{14}\right)}{64 \pi \, \vs c_{14} \GAE} \int r^2 \, \epsilon^i{}_{jk} \left[x^j \left(\pd^k F\right)\left(\pd_\tau F\right)\right] d\Omega\,.
	\end{align}
\end{subequations}
Notice that the tensor term in Eq.~\eqref{eq:AngularMomentomVector} matches that given in GR except for the prefactor in front of the integral.  The largest difference between these expressions and those in GR is the presence of the vector and scalar modes, which contribute to the total loss of angular momentum of the system. Even if one sets $c_{+} = 0$ to satisfy the tensor propagation bounds of~\cite{GBM:2017lvd}, the vector and scalar contributions still remain. 

For completeness, we show the rate of energy loss in terms of our fields\footnote{See~\cite{Foster:2006az,Yagi:2013ava,Saffer:2017ywl} for a calculation on how $\dot{E}$ was obtained.}
\begin{subequations}
	\label{eq:Edot1}
	\begin{align}
		\dot{E}_{(T)} &= -\frac{1}{32\pi \, \vt \GAE} \int r^2 \left(\partial_\tau\phi^{\TT}_{ij}\right) \left(\partial_\tau\phi_{\TT}^{ij}\right) d\Omega \,,\\
		\dot{E}_{(V)} &= -\frac{(1+c_+)(2c_1-c_+c_-)}{16\pi \, \vv \GAE} \int r^2 \left(\partial_\tau \nu_i\right) \left(\partial_\tau \nu^i\right) d\Omega \,, \\
		\dot{E}_{(S)} &= -\frac{(2-c_{14})}{64\pi \, \vs \,c_{14} \, \GAE} \int r^2 \left(\pd_\tau F\right)\left(\pd_\tau F\right) d\Omega \,.
	\end{align}
\end{subequations}
It is a curious occurrence that all factors of our coupling constants are identical in the prefactors for both $\dot{L}^i$ and $\dot{E}$ for their respective modes.  We will make further use of Eq.~\eqref{eq:Edot1} below.

\subsection{Specialization to an N-Body System}
\label{sec:BinarySection}

The solution to the \AE{} field equations for a PN binary system were first calculated in~\cite{Foster:2006az} and found to be
\begin{align}
\label{eq:TensorExpansion}
	\phi^{\TT}_{ij} &= \frac{2\,G_{\AE{}}}{r} \, \ddot{Q}^\TT_{ij}\left(t-\frac{r}{ \vt }\right)\,,
	\\
\label{eq:VectorExpansion}
	\nu^i &= -\frac{2 G_{\AE{}}}{\left(2\,c_1 - c_+c_-\right)r} \nonumber \\
	& \times \left[\frac{1}{v_V}\left(\frac{c_+}{1-c_+} \, \ddot{Q}^i{}_{j}+\ddot{\mathcal{Q}}^i{}_{j} + \mathcal{V}^i{}_{j}\right)N^j - 2 \Sigma^i\right]^{\mbox{\tiny T}}\,,
	\\
    F &= \frac{2\,c_{14} \, G_{\AE{}}}{\left(2 - c_{14}\right)r} \left[3\left(Z-1\right)N^i N^j\ddot{Q}_{ij} + Z\,\ddot{I}^k{}_k \right. \nonumber \\
	& \left. - \frac{1}{c_{14}\,\vs}\left(\ddot{\mathcal{Q}}_{ij} + \frac{1}{3} \,\delta_{ij} \, \ddot{\mathcal{I}}^k{}_k\right)N^iN^j + \frac{2}{c_{14} \, \vs} N^i \,\Sigma_i\right]\,,
\end{align}
where over-head dots stand for partial time derivatives, the superscript T and TT stand for the transverse and the transverse-traceless part (relative to the direction of propagation) of a given object, and we have defined the short-hands
\begin{align}
	Q_{ij} &= I_{ij} - \frac{1}{3}\,\delta_{ij} \, I^k{}_k\,,
	\\
	\mathcal{Q}_{ij} &= \mathcal{I}_{ij} - \frac{1}{3}\,\delta_{ij} \, \mathcal{I}^k{}_k\,.
\end{align}
These shorthands depend on the multipole moments of a system of $A$ bodies
\begin{align}
	I^{ij} &= \sum_A \, m_A x_A^i x^j_A\,,
	\\
	\mathcal{I}^{ij} &= \sum_A \sigma_A\, \tilde{m}_A x_A^i x^j_A\,,
	\\
	\Sigma^i &= - \sum_A \sigma_A \tilde{m}_A \, v^i_A\,,
	\\
	\mathcal{V}^{ij} &= 2 \sum_A \sigma_A \tilde{m}_A \dot{v}_A^{[i}x_A^{j]}\,,
\end{align}
where $x_{A}^{i}$ labels their spatial trajectories and $v^i_A = \dot{x}^i_A$ their orbital velocities,
and we have defined the constants~\cite{Foster:2005dk}
\begin{align}
	Z &= \frac{2\left(1-c_+\right)\left(\alpha_1 - 2\, \alpha_2\right)}{3 \left(2c_+ - c_{14}\right)}
	\\
	\alpha_1 &= - \frac{8\left(c_1 \, c_4 +c_3^2\right)}{2c_1-c_+c_-} \,,
	\\
	\alpha_2 &= \frac{\alpha_1}{2} - \frac{\left(c_1+2c_3-c_4\right)\left(2c_2+c_{123}+c_{14}\right)}{c_{123}\left(2-c_{14}\right)}\,.
\end{align}
All expressions here are to be evaluated at the retarded time $t-r/v$, where $v$ is to be replaced with the propagation velocity of the corresponding mode. 

With this at hand, we may utilize the Brill-Hartle averaging scheme~\cite{PhysRev.135.B271} and write the angular momentum loss rate as 
\begin{widetext}
\begin{align}
\dot{L}^{i}_{(T)} &= -\frac{2 \, \GAE}{5 \, \vt} \, \epsilon^i{}_{jk}\avg{\ddot{Q}^{ja} \, \dddot{Q}^k{}_a}\,.
	\\
\label{eq:LdotVector}
	\dot{L}^{i}_{(V)} &= -\frac{4 \, c_{14} \, c_+^2 \, \GAE}{5 \left(2\,c_1-c_+c_-\right)^2 \vv} \epsilon^i{}_{jk} \avg{\ddot{Q}^{ja}\dddot{Q}^k{}_a} - \frac{2 \, c_+\,\GAE}{5 \left(2c_1 - c_+c_-\right)\vv^3} \epsilon^i{}_{jk} \avg{ \ddot{\mathcal{Q}}^{ja}\dddot{Q}^k{}_a +\ddot{Q}^{ja}\dddot{\mathcal{Q}}^k{}_a } \nonumber \\
	& - \frac{2 \left(1-c_+\right)\GAE}{5\left(2c_1 - c_+c_-\right) \vv^3} \epsilon^i{}_{jk} \avg{\ddot{\mathcal{Q}}^{ja}\dddot{\mathcal{Q}}^k{}_a} - \frac{2 \left(1-c_+\right)\GAE}{3\left(2c_1 - c_+c_-\right) \vv^3} \epsilon^i{}_{jk} \avg{ \mathcal{V}^{ja}\dot{\mathcal{V}}^k{}_a} - \frac{8 \left(1-c_+\right) \GAE}{3 \left(2c_1-c_+c_-\right) \vv} \epsilon^i{}_{jk} \avg{\Sigma^j \dot{\Sigma}^k} \,, 
	\\
	\label{eq:LdotScalar}
	\dot{L}^i_{(S)} &= -\frac{3 \, c_{14} \, \GAE}{5\left(2-c_{14}\right)\vs}\left(Z-1\right)^2 \epsilon^i{}_{jk} \avg{\ddot{Q}^{ja}\dddot{Q}^k{}_a} + \frac{2 \, \GAE}{5\left(2-c_{14}\right) \vs^3} \left(Z-1\right)\epsilon^i{}_{jk}\avg{ \ddot{\mathcal{Q}}^{ja}\dddot{Q}^k{}_a + \ddot{Q}^{ja} \dddot{\mathcal{Q}}^k{}_a} \nonumber \\
	& - \frac{4 \, \GAE}{15 \left(2-c_{14}\right) c_{14} \, \vs^5} \epsilon^i{}_{jk} \avg{\ddot{\mathcal{Q}}^{ja} \dddot{\mathcal{Q}}^k{}_a} - \frac{2 \, \GAE}{3 \left(2-c_{14}\right)c_{14} \, \vs^3} \epsilon^i{}_{jk} \avg{\Sigma^j \dot{\Sigma}^k}\,.
	\end{align}
The above expressions can be combined and simplified to write the total angular momentum loss as follows 
\begin{align}
\label{eq:LdotFinal}
	\dot{L}^i &= -\GAE \, \epsilon^i{}_{jk} \avg{\frac{2\, \mathcal{A}_1 }{5}\, \ddot{Q}^{ja}\dddot{Q}^k{}_a + \frac{\mathcal{A}_2}{5}\left( \ddot{\mathcal{Q}}^{ja}\dddot{Q}^k{}_a + \ddot{Q}^{ja} \dddot{\mathcal{Q}}^k{}_a\right)+ \frac{2 \, \mathcal{A}_3}{5} \, \ddot{\mathcal{Q}}^{ja} \dddot{\mathcal{Q}}^k{}_a + \mathcal{C} \, \Sigma^k \, \dot{\Sigma}^k +2 \, \mathcal{D} \, \mathcal{V}^{ja}\dot{\mathcal{V}}^k{}_a}\,,
\end{align}
\end{widetext}
where we have introduced the new constants
\begin{subequations}
	\label{eq:constants}
	\begin{align}
		\mathcal{A}_1 &\equiv \frac{1}{\vt} + \frac{2 \, c_{14} \, c_+^2}{\left(2c_1 - c_+c_-\right) \vv}+\frac{3\,c_{14}\left(Z-1\right)^2}{2\left(2-c_{14}\right)\vs}\,, \\
		\mathcal{A}_2 &\equiv \frac{2 \, c_+}{\left(2c_1-c_+c_-\right)\vv^3} - \frac{2\left(Z-1\right)}{\left(2-c_{14}\right)\vs^3}\,, \\
		\mathcal{A}_3 &\equiv \frac{1}{2\, c_{14} \, \vv^5} + \frac{2}{3\,c_{14}\left(2-c_{14}\right)\vs^5}\,, \\
		\mathcal{C} &\equiv \frac{4}{3\,c_{14} (2-c_{14})\vs^3} + \frac{4}{3\,c_{14}\,\vv^3}\,, \\
		\mathcal{D} &\equiv \frac{1}{6 \, c_{14} \, \vv^5}\,.
	\end{align}
\end{subequations}
The first term in the $\mathcal{A}_1$ quantity represents the GR term (with the modified propagation velocity).  All other terms lead to deviations away from GR due to contributions of the \ae{}ther field.  As with the discussion above, in the limit $c_+ \to 0$ the contributions of the additional modes of propagation persist.  

For completeness, we also present here the energy loss for an N-Body system, computed e.g. in~\cite{Yagi:2013ava} 
\begin{align}
\label{eq:Edot2}
	\dot{E} &= - \GAE\avg{\frac{\mathcal{A}_1}{5}\, \dddot{Q}^{ij} \dddot{Q}_{ij} + \frac{\mathcal{A}_2}{5} \, \dddot{\mathcal{Q}}^{ij}\dddot{Q}_{ij} + \frac{\mathcal{A}_3}{5} \, \dddot{\mathcal{Q}}^{ij} \dddot{\mathcal{Q}}_{ij} \right. \nonumber \\
	& \left. + \, \mathcal{B}_1 \dddot{I} \dddot{I} + \mathcal{B}_2 \dddot{\mathcal{I}} \dddot{I} + \mathcal{B}_3 \dddot{\mathcal{I}}\dddot{\mathcal{I}} + \mathcal{C} \dot{\Sigma}^i \dot{\Sigma}_i + \mathcal{D} \dot{\mathcal{V}}^{ij}\dot{\mathcal{V}}_{ij}}\,,
\end{align}
where 
\begin{subequations}
	\begin{align}
		\mathcal{B}_1 &\equiv \frac{c_{14} \, Z^2}{4\,(2-c_{14})\, \vs}\,, \\
		\mathcal{B}_2 &\equiv \frac{Z}{3\,(c_{14}-2) \vs^3}\,,\\
		\mathcal{B}_3 &\equiv \frac{1}{9\,c_{14}\,(2-c_{14})\vs^5}\,.
	\end{align}
\end{subequations}
This result is important for use in the following section.  As with the $\dot{L}^i$ term, the deviations away from the GR result for $\dot{E}$ are due to the presence of the alternative modes whose velocities are unbounded.

\section{Binary Dynamics in \AE{}-theory}
\label{sec:BinarySection-AE}

\begin{figure}
	\centering
	\vspace{0.3cm}
	\includegraphics[width=0.3\textwidth]{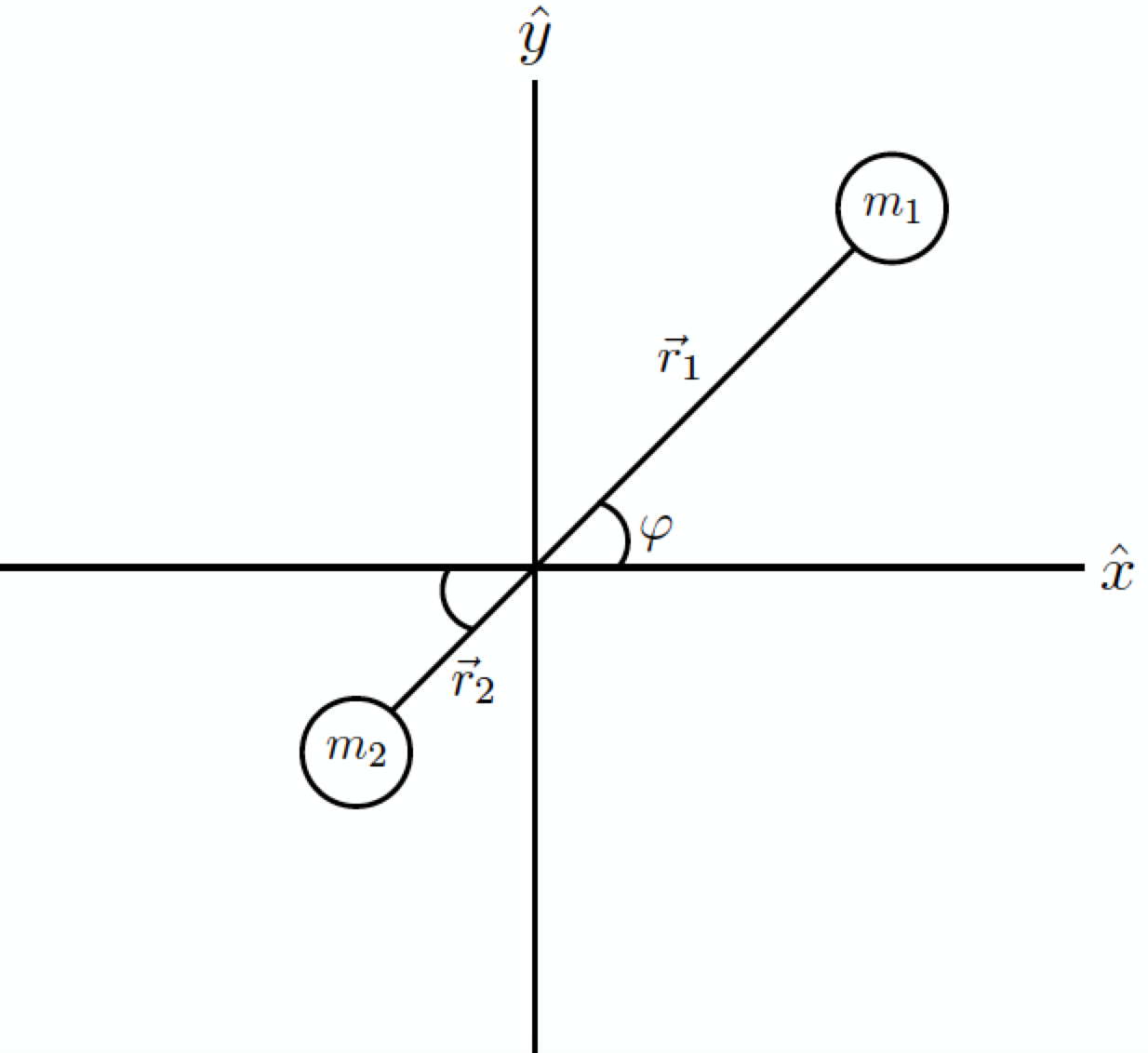}
	\caption{Example of Binary System Used}
	\label{fig:binary}
\end{figure}

We now specialize the discussion further by considering a non-spinning binary system with active gravitational masses $m_{1}$ and $m_{2}$ in an elliptical orbit of separation $\vec{r}_{12} = \vec{r}_{1} - \vec{r}_{2}$, as shown schematically in Fig.~\ref{fig:binary}.  Observe that the orbital plane is chosen here to coincide with the $x$-$y$ plane, so that the orbital angular momentum is along the $z$-axis.  

Neglecting radiation-reaction, such an orbit can still be described through Keplerian ellipses in \AE{}-theory~\cite{Foster:2007gr,Yagi:2013ava}.
\begin{align}
		r_{12} &= \frac{a(1-e^2)}{1 + e \cos(\varphi)}\,, 
\end{align}
where the azimuthal angle evolves according to
\begin{align}
	\dot{\varphi} &= \frac{1}{r_{12}^2}\sqrt{\mathcal{G}\,M\,a\,(1-e^2)}\,,
\end{align}
and where $a$ is the semi-major axis, $e$ is the orbital eccentricity, $M=m_1+m_2$ is the total mass and we have defined 
\begin{align}
	\mathcal{G} &\equiv \frac{G_N}{(1+\sigma_1)(1+\sigma_2)}\,.
\end{align}
Without radiation-reaction, the semi-major axis $a$ and the orbital eccentricity $e$ are constants, while $\phi$ is a function of time. The binary's energy and angular momentum are then also constant, and related to the semi-major axis and the eccentricity via
\begin{subequations}
	\label{eq:EandL}
	\begin{align}
		E &= - \frac{\mathcal{G} \, m_1 \,m_2}{2a}\,,\\
		L^2 &= \frac{\mathcal{G} \, m_1^2 \, m_2^2}{M} \, a(1-e^2)\,.
	\end{align}
\end{subequations}

Once we include radiation-reaction, the energy and angular momentum are no longer constant, as we found in the previous section. Instead, for a binary system,  Eqs.~\eqref{eq:Edot2} and~\eqref{eq:LdotFinal} are given by
\begin{align}
\label{eq:Edot3}
	\dot{E} &= - \frac{\mathcal{C} \, \GAE^2 \, \mathcal{G}^2\left(2-e^2-e^4\right)S_-^2 \, M^4\, \eta^2 }{2\left(1-e^2\right)^{7/2}a^4} \nonumber \\
	&- \frac{\GAE \, \mathcal{G}^3 \, M^5 \, \eta^2}{30 \left(1-e^2\right)^{7/2}a^5} \nonumber \\
	&\times \left[2\left(96+292e^2+37e^4\right)\left(\mathcal{A}_1 + \mathcal{S}\mathcal{A}_2 + \mathcal{S}^2\mathcal{A}_3\right) \nonumber \right. \\
	&\left. + 15e^2\left(4+e^2\right)\left(\mathcal{B}_1 + \mathcal{S}\mathcal{B}_2 + \mathcal{S}^2\mathcal{B}_3\right)\right]\,,
\end{align}
and
\begin{align}
\label{eq:Ldot3}
	\dot{L}^i &= -\frac{\mathcal{C} \, \GAE \, \mathcal{G}^{3/2} \, S_-^2 \, M^{7/2} \, \eta^2}{\left(1-e^2\right) a^{5/2}} \nonumber \\
	&- \frac{4\left(8+7e^2\right)\GAE \, \mathcal{G}^{5/2} \, M^{9/2}\, \eta^2}{5\left(1-e^2\right)^2 a^{7/2}}\left(\mathcal{A}_1 + \mathcal{S}\mathcal{A}_2 + \mathcal{S}^2\mathcal{A}_3\right)
\end{align}
where we introduced the symmetric mass ratio $\eta=m_1m_2/M^2$, the mass-weighted symmetric sensitivity $\mathcal{S}=m_1s_2/M+m_2s_1/M$, and the anti-symmetric sensitivity $S_-=s_1-s_2$.

The above expressions shown in Eqs.~\eqref{eq:Edot3} and~\eqref{eq:Ldot3} do not contain angle-brackets because we have explicitly carried out the orbit averaging. By the latter, we mean specifically 
\begin{equation}
\label{eq:averaging1}
	\avg{X(t)} = \frac{1}{P}\int_0^{P} X(t)\,dt\,,
\end{equation}
for any quantity $X(t)$ with $P$ the orbital period. This integral can of course be rewritten in terms of the orbital angle ($\varphi$ in Fig.~\ref{fig:binary}), as
\begin{equation}
	\avg{X(\varphi)} = \frac{\left(1-e^2\right)^{3/2}}{2\pi} \int_0^{2\pi} \frac{X(\varphi)}{\left(1+e\cos(\varphi)\right)^2}\,d \varphi\,.
\end{equation}
where we have used 
\begin{equation}
\frac{d \varphi}{d t} = \sqrt{\frac{\GAE \, M}{a^3}} \left(1-e^2\right)^{-3/2} \left(1+e\cos\left(\varphi\right)\right)^2\,.
\end{equation}

We can now combine the above equations with the time derivative of Eq.~\eqref{eq:EandL} to find how the semi-major axis and the orbital eccentricity evolve. We find that
\begin{widetext}
\begin{align}
\label{eq:adot}
	\dot{a} &= -\frac{\mathcal{C}\,\GAE \, \mathcal{G} \, S_-^2 \, M^2\,\eta}{a^2} \, \text{h}_1(e) 
	- \frac{\GAE \, \mathcal{G}^2\,M^3\,\eta}{15\,a^3}\left[2\left(\mathcal{A}_1 + \mathcal{S}\mathcal{A}_2 + \mathcal{S}^2\mathcal{A}_3\right) \text{f}_1(e) 
	+ 15\left(\mathcal{B}_1 + \mathcal{S}\mathcal{B}_2 + \mathcal{S}^2\mathcal{B}_3\right) \text{g}_1(e)\right]\,,
	\\
	\label{eq:edot}
	\dot{e} &= -\frac{\mathcal{C} \,\GAE \, \mathcal{G} \, S_-^2 M^2 \, \eta}{a^3}\,\text{h}_2(e) - \frac{\GAE \, \mathcal{G}^2 \, M^3 \, \eta}{30 \, a^4} \left[2\left(\mathcal{A}_1 + \mathcal{S}\mathcal{A}_2 + \mathcal{S}^2\mathcal{A}_3\right)\text{f}_2(e) + 15\left(\mathcal{B}_1 + \mathcal{S}\mathcal{B}_2 + \mathcal{S}^2\mathcal{B}_3\right) \text{g}_2(e)\right]\,.
\end{align}
\end{widetext}
The above expressions depend on certain enhancement factors that are only functions of eccentricity, namely
\begin{subequations}
	\begin{align}
		\text{f}_1(e) &\equiv \frac{96 + 292 e^2 + 37 e^4}{(1 - e^2)^{7/2}}\,,\\
		\text{g}_1(e) &\equiv \frac{e^2 (4 + e^2)}{(1 - e^2)^{7/2}}\,,\\
		\text{h}_1(e) &\equiv \frac{2+e^2}{(1-e^2)^{5/2}}\,, \\
		\text{f}_2(e) &\equiv  \frac{e\left(304 + 121\,e^2\right)}{(1-e^2)^{5/2}}\,,\\
		\text{g}_2(e) &\equiv \frac{e\left(4+e^2\right)}{(1-e^2)^{5/2}}\,, \\
		\text{h}_2(e) &\equiv \frac{3\,e}{2(1-e^2)^{3/2}} \,,\\
	\end{align}
\end{subequations}
The evolution of the semi-major axis and the orbital eccentricity is clearly faster than in GR due to the dipole term proportional to ${\cal{C}}$.  As was shown earlier, $\mathcal{C}$ is produced by terms that come from the vector and scalar modes of propagation. Therefore, it is these additional propagation modes that are responsible for the contribution to the dipole term in the decay rates, and the latter is thus not affected by bounds on the propagation speed of the tensor mode.

Figure~\ref{fig:changes} shows the temporal evolution of the semi-major axis and the orbital eccentricity for a neutron star binary in the inspiral. This evolution was obtained by numerically solving Eqs.~\eqref{eq:adot} and~\eqref{eq:edot} for a binary with masses $m_{1}=1.365 M_{\odot}$ and $m_{2}=2.040 M_{\odot}$. The sensitivities are modeled through a weak-field expansion of~\cite{Foster:2006az} with stellar radii $R_{1}=12.214$km and $R_{2}=11.966$km and are given by
\begin{equation}
	s_A = \left(\alpha_1 - \frac{2}{3} \, \alpha_2\right)\frac{m_A}{R_A}\,.
\end{equation}
This approximation suffices to roughly model the temporal evolution of the semi-major axis and the orbital eccentricity. A proper GW model in \AE{}-theory, however, ought to employ the strong-field representation of the sensitivities found in~\cite{Yagi:2013ava}. 

\section{Conclusion and Future}
\label{sec:Conclusions}

We have here calculated the rate of change of the orbital angular momentum due to GW emission in \AE{}-theory, accounting for all modes of propagation. This calculation, together with the energy flux of~\cite{Foster:2006az,Yagi:2013ava}, allowed us to compute the rate of change of a binary's semi-major axis and its orbital eccentricity. We found that these quantities decay much more rapidly than in GR due to the emission of dipole radiation by the vector and scalar modes of the theory. These modifications persist even when one forces the tensor modes to propagate at the speed of light, thus by-passing the recent GW bounds of~\cite{GBM:2017lvd}. 

One could use these results in a variety of ways. One possibility is to construct a leading order GR deviation model for the GWs emitted in the inspiral of eccentric compact binaries. This model could be based on the recent work of~\cite{Moore:2018kvz}, which built a model in GR that is valid for binaries with arbitrary eccentricity. Alternatively, one could work in a small-eccentricity expansion, as suggested in~\cite{Yunes:2009yz}, or in a large-eccentricity expansion, as suggested in~\cite{Loutrel:2017fgu,Loutrel:2014vja}. Such a model could then be used to inspiral a model-independent parameterization of modified GW models, as done for quasi-circular inspirals through the parameterized post-Einsteinian framework~\cite{Yunes:2009ke}. With this at hand, one could use future GW observations by advanced LIGO, or by third generation GW detectors or space-based detectors~\cite{Audley:2017drz}, to place interesting constraints on \AE{}-theory.    

Another possibility for future work is to revisit the strong-field sensitivity calculations of~\cite{Yagi:2013ava} and repeat the computation of binary pulsar constraints in light of the recent tensor propagation speed constraints of~\cite{GBM:2017lvd}. The results on the \AE{} constraints found in~\cite{Yagi:2013ava} are not valid any longer because that work made certain assumptions regarding the magnitude of the coupling constants, which is violated by the constraints in~\cite{GBM:2017lvd}. Once this work is repeated, however, one could redo the computation of binary pulsar constraints using not only binaries with very small eccentricity, but also much more eccentric ones. Such an extension would require the use of the angular momentum flux calculated in this paper.   
 
\section{Acknowledgments}
\label{ackno}

N.Y and A.S.~acknowledge support through the NSF CAREER grant PHY-1250636 and NASA grants NNX16AB98G and 80NSSC17M0041.

\bibliography{LdotBib}
\end{document}